\documentclass[a4paper,fleqn,usenatbib]{mnras}

\usepackage{mathptmx}

\usepackage[T1]{fontenc}
\usepackage{ae,aecompl}

\usepackage{graphicx}	
\usepackage{amsmath}	
\usepackage{amssymb}	
\usepackage{nccmath}

\title[Lensed quasars in \textit{Gaia}]{Gravitationally Lensed Quasars in \textit{Gaia}: I. Resolving Small-Separation Lenses}

\author[C. A. Lemon et al.]{Cameron A. Lemon$^{1, 2}$\thanks{E-mail: cl522@ast.cam.ac.uk}, 
Matthew W. Auger$^{1}$,
Richard G. McMahon$^{1, 2}$,
\newauthor
Sergey E. Koposov$^{3}$
\\
$^{1}$Institute of Astronomy, University of Cambridge, Madingley Road, Cambridge CB3 0HA, UK\\
$^{2}$Kavli  Institute  for  Cosmology,  University  of  Cambridge,  Madingley Road, Cambridge CB3 0HA, UK\\
$^{3}$McWilliams Center for Cosmology, Department of Physics, Carnegie Mellon University, 5000 Forbes Avenue, Pittsburgh, PA 15213, USA
}

\date{Accepted XXX. Received YYY; in original form ZZZ}

\pubyear{2015}

\begin{document}
\label{firstpage}
\pagerange{\pageref{firstpage}--\pageref{lastpage}}
\maketitle

\begin{abstract}

\textit{Gaia}'s exceptional resolution (FWHM $\sim$ 0.1$\arcsec$) allows identification and cataloguing of the multiple images of gravitationally lensed quasars. We investigate a sample of 49 known lensed quasars in the SDSS footprint, with image separations less than 2$\arcsec$, and find that 8 are detected with multiple components in the first \textit{Gaia} data release. In the case of the 41 single \textit{Gaia} detections, we generally are able to distinguish these lensed quasars from single quasars when comparing \textit{Gaia} flux and position measurements to those of Pan-STARRS and SDSS. This is because the multiple images of these lensed quasars are typically blended in ground-based imaging and therefore the total flux and a flux-weighted centroid are measured, which can differ significantly from the fluxes and centroids of the individual components detected by \textit{Gaia}. We compare the fluxes through an empirical fit of Pan-STARRS \textit{griz} photometry to the wide optical \textit{Gaia} bandpass values using a sample of isolated quasars. The positional offsets are calculated from a recalibrated astrometric SDSS catalogue. Applying flux and centroid difference criteria to spectroscopically confirmed quasars, we discover 4 new sub-arcsecond-separation lensed quasar candidates which have two distinct components of similar colour in archival CFHT or HSC data. Our method based on single \textit{Gaia} detections can be used to identify the $\sim$ 1400 lensed quasars with image separation above 0.5$\arcsec$, expected to have only one image bright enough to be detected by \textit{Gaia}.

\end{abstract}




\section{Introduction}
Gravitationally lensed quasars are powerful tools to study extragalactic and cosmological phenomena. As with all lenses, the unique geometry can be used for detailed study of both the sources \citep[e.g.][]{stacey2017, ding2017} and also the lensing galaxies \citep[e.g.][]{oguri2014, birrer2017}. Since quasars are variable sources, the time delays between the lightcurves of the individual lensed images can be measured \citep[e.g.][]{eulaers2013}. These time delays depend on the different path lengths from source to observer and the gravitational potential along these paths and thus can be used to determine the Hubble constant \citep{refsdal1966}, recently to better than 4 percent precision \citep{bonvin2016}.

Each lightcurve is imprinted with further variations due to microlensing by stars in the lensing galaxy, making it possible to probe the structure of the source quasar accretion disk \citep[e.g.][]{bate2008, motta2017} and dark matter fractions in the lensing galaxy \citep{bate2011, schechter2014, jimenez2015}. Though more than 150 lensed quasars are known to date, only a subset can be used to enable the above science. For example, microlensing events may be most noticeable in systems with low-redshift lenses, leading to smaller microlensing timescales \mbox{\citep[e.g.][]{mosquera2011}}.

The methods developed in this paper are aimed at finding small-separation lensed quasars, with image separations typically less than 1$\arcsec$. These lenses, though more difficult to find, are more common than wider-separation systems and can be used to probe either higher-redshift/lower-mass lensing galaxies and to understand mass distributions in the cores of galaxies \citep[e.g.][]{trott2002}. Furthermore, forming a complete sample of lensed quasars at lower separations will yield better constraints on the cosmological mass density parameter and evolution of the lensing galaxy population  \citep{oguri2012, finet2016}.

Dedicated follow-up programs to determine statistical samples of lensed quasars have been based on radio surveys \citep[e.g. Cosmic Lens All-Sky Survey (CLASS),][]{myers1995} or wide-field optical surveys \citep[e.g. Sloan Digital Sky Survey (SDSS) Quasar Lens Search (SQLS),][]{oguri2006}. Each survey has its own advantages and disadvantages; radio surveys like CLASS were only able to target radio-loud lensed sources but discovered smaller image separation objects, and SQLS had multi-band optical imaging of a large fraction of the sky, but was incomplete towards the smallest-separation lenses due to the median seeing of 1.4$\arcsec$ in SDSS. Accordingly the median image separation of the CLASS and SQLS lenses are 0.8$\arcsec$ and 1.65$\arcsec$ respectively.

Many techniques have been suggested in the literature for finding such rare systems in these vast datasets. Each method is tailored to different types of data, e.g. multi-band colour classification to find quasar-like systems which are extended or deblended into multiple components \citep{ostrovski2016}. Other techniques rely on the time domain to reveal neighbouring variable objects \citep{kochanek2006} or search for time delays through cross-correlations \citep{pindor2005}.  However, many current wide-field surveys lack the necessary number of epochs to apply these methods. In many cases the lenses must be found through single-epoch imaging and in the case of some searches start from spectroscopically confirmed quasars.

With the advent of many wide-field optical surveys covering the whole sky, \citet{oguri2010} (hereafter OM10) predicted lensed quasar numbers expected in such surveys. Crucially, these numbers rely on all lensed quasars being detected at separations above two thirds of the point spread function (PSF) full width half-maximum (FWHM) down to the limiting magnitudes of the survey. This is a difficult task without many epochs of imaging, and so further techniques must be developed to meet this goal. It is important to note that, in agreement with these predictions, datasets like SDSS are not depleted of bright lensed quasars since searches like SQLS are biased towards certain source redshifts and quasar selection criteria  \citep{richards2002}. SQLS also mainly targeted the candidate systems with image separation $\gtrsim$ 1$\arcsec$. Indeed lensed quasars are being found in SDSS photometric data alone \citep[Ostrovski et al. in prep., ][]{agnello2017} without spectroscopic selection.

\begin{figure}
\begin{tabular}{ll}
\includegraphics[scale=0.31415]{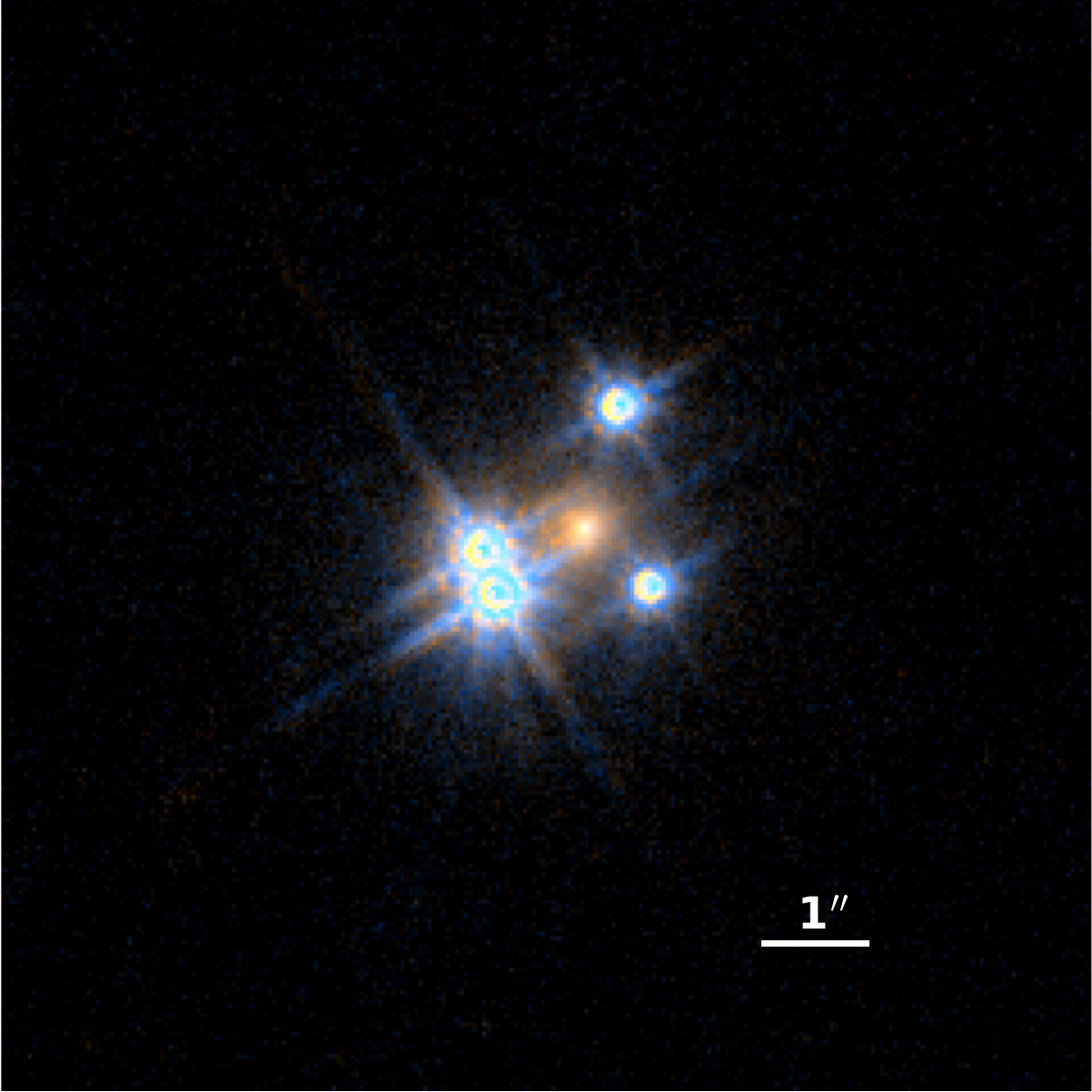}
&
\includegraphics[scale=0.31415]{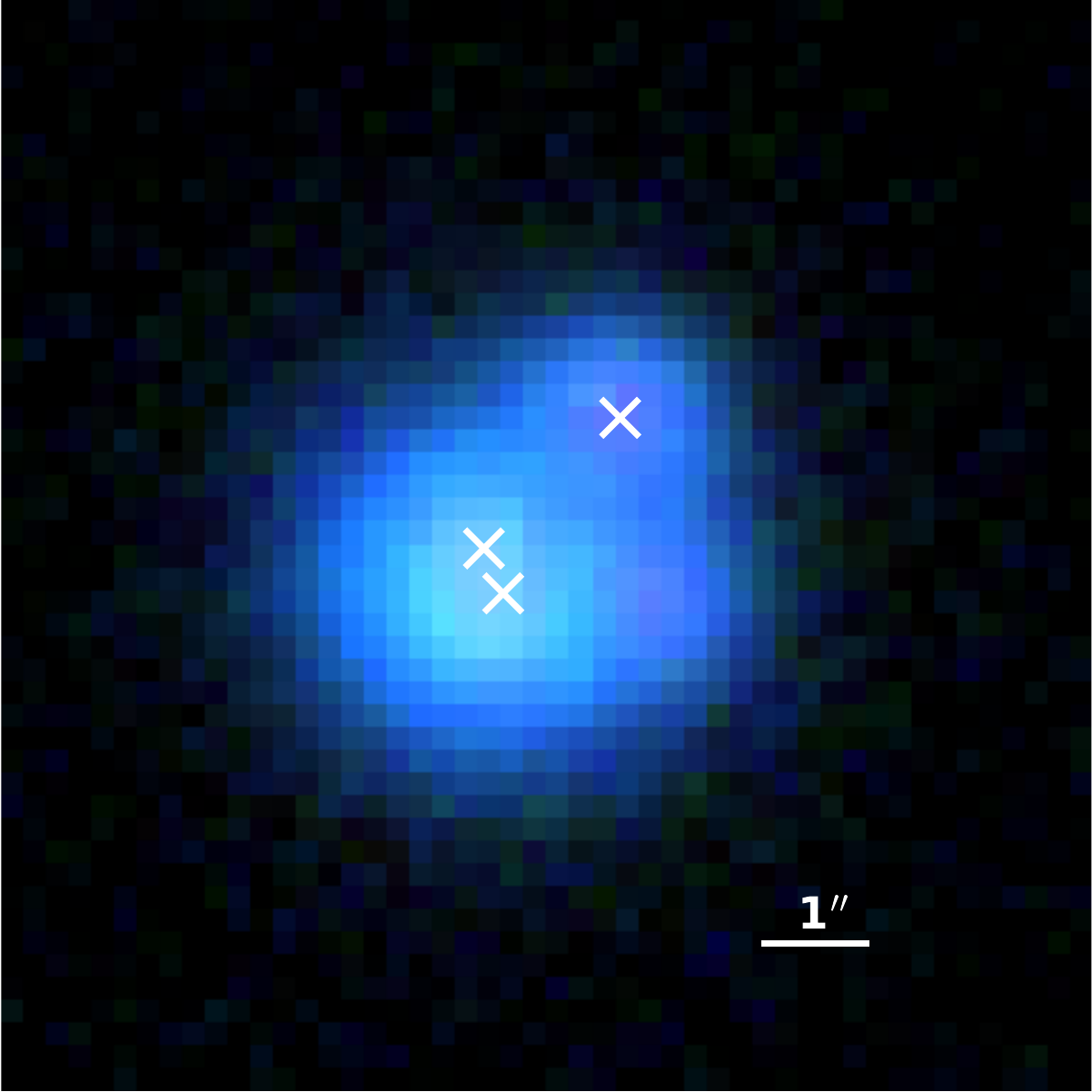}
\end{tabular}
\caption{PG1115+080 as seen in HST (left) and Pan-STARRS (right). \textit{Gaia}'s excellent resolution allows it to identify the three brightest quasar images, including both images in the merging pair, separated by 0.5$\arcsec$. \textit{Gaia} detections are marked with white crosses.}
\label{fig:PG1115}
\end{figure}

In this paper we show that the recent \textit{Gaia} data release \citep{lindgren2016} is a very useful tool to overcome the difficulties of finding lensed quasars in ground-based surveys alone, based mainly on its excellent angular resolution and full-sky coverage. Figure \ref{fig:PG1115} demonstrates how \textit{Gaia} is able to detect three of the four quasar images of PG1115+080 \citep{weymann1980}. The methods we develop are useful to select any optical binary. This includes stellar binaries which are key to understanding star formation and evolutionary models \citep[e.g.][]{chabrier2003, liu2008} and also quasar pairs, including binary quasars and projected pairs \citep[e.g.][]{hennawi2006}. The brightest close-separation binary companions are often found in ground-based imaging data, using similar methods to quasar lens searches, such as looking for extended systems inconsistent with a single PSF \citep{deacon2017}. However this is an incomplete method at small separations, with pipeline PSF models unknowingly using these binaries to determine the local PSF, thus affecting the PSF magnitudes.

In Section 2 of this paper we outline the most important features of the \textit{Gaia} satellite and how its source detection relates to quasars and lensed quasars. In section 3 we consider this in the context of a sample of known lensed quasars with small image separations. In Sections 4 and 5 we present a technique to use \textit{Gaia} to determine lens candidates and apply this technique to SDSS spectroscopically confirmed quasars, presenting  potential small-separation lens candidates. We summarise our findings in Section 6.

\section{\textit{Gaia}} 
\textit{Gaia} is a space-based mission mapping the stars of the Milky Way with unprecedented astrometric precision. It is able to detect bright quasars, making it a useful tool for lensed quasar searches. In this section we describe the basic \textit{Gaia} mission and satellite details to explain the cataloguing of lensed quasars in \textit{Gaia}'s first data release (hereafter GDR1).

\subsection{Overview of the \textit{Gaia} mission and satellite}
The \textit{Gaia} satellite \citep{prusti2016} was launched on 19 December 2013, with GDR1 consisting of observations from the first 14 months (25 July 2014 to 16 September 2015) of the nominal 60 month mission. \textit{Gaia} consists of two identical telescopes, each with a rectangular aperture of 1.45m$\times$0.45m, that simultaneously point in directions separated by 106.5 degrees with beams folded to a common focal plane. As a result of the asymmetric focal plane with ratio 3:1, the PSF is also asymmetric with the same ratio and has a measured median FWHM of 103 mas \citep{fabricius2016} in the scanning axis direction. \textit{Gaia} is located at L2 with a rotational sky-scanning orbit of period 6 hours and an orbital precession period of 63 days. Over the course of \textit{Gaia}'s 5 year mission it will measure each source $\sim$ 70 times.

\subsection{\textit{Gaia} focal plane}
The \textit{Gaia} focal plane consists of 106 CCDs with 4500 pixels in the along-scan (AL) direction and 1966 pixels in the across-scan (AC) direction. Each pixel is rectangular in the ratio 1:3 similar to the PSF major and minor axes with size 10$\times$30 microns corresponding to 59$\times$177 mas on the sky, i.e. $\sim$ 2 pixels Nyquist sampling of the PSF FWHM. There are 14 sky-mapper (SM) CCDs and 62 astrometric field (AF) CCDs aligned in 7 rows in the AL direction and 9 columns in the AC direction with the middle CCD of the 9th AF column (AF9) assigned to one of the two focus wave front sensors (WFS). The SM CCDs are in two columns (SM1 and SM2) with baffling such that each SM can only view a single \textit{Gaia} telescope whereas the AF CCDs view the two fields simultaneously. The integration time per CCD is 4.42 seconds but it is impossible to download all the pixels to Earth and hence astrometric measurements are made via windowed regions for sources detected in the SM CCDs.

\begin{figure}
	\includegraphics[trim={2.5cm 0.2cm 2.5cm 1cm},clip, width=\columnwidth]{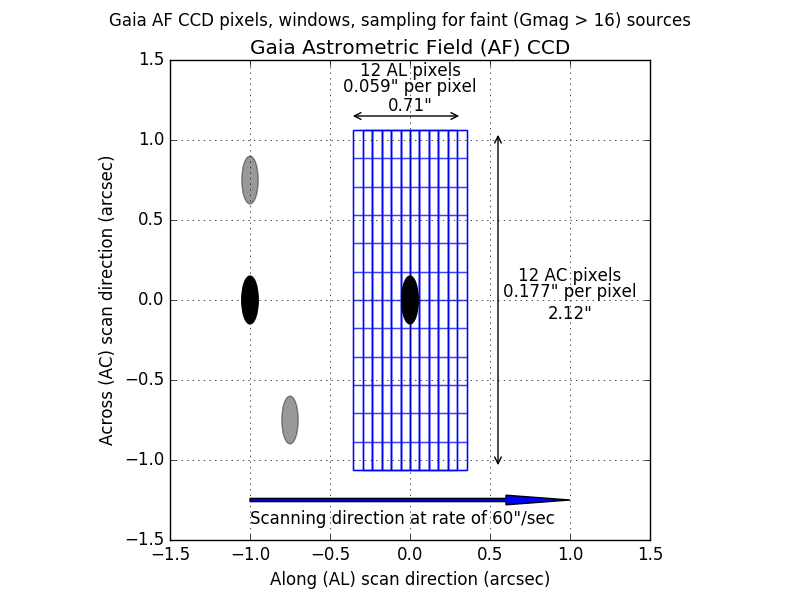}
    \caption{A schematic of the sampling of a faint (G>16) source in a \textit{Gaia} astrometric field CCD. The ellipses represent the images of point sources on the focal plane.}
    \label{fig:Gaia_af}
\end{figure}

For sources brighter than G=13, 2D (AF) windows of 12 AC pixels (2.12$\arcsec$) and 18 AL pixels (1.06$\arcsec$) are transmitted to ground \citep{carrasco2016}. For all fainter sources (G>13) the windows are binned in the AC direction during readout to produce a 1D sample. For the faint sources brighter than G=16 (i.e. 13<G<16) there are 18 AL samples (Long 1D). For sources fainter than G=16 there are 12 AL samples (Short 1D) with total length of 0.71$\arcsec$. Figure \ref{fig:Gaia_af} shows the 12$\times$12 pixel 2D window for a faint source, for which a 1D data stream of 12 samples is downloaded with a sampling resolution of 0.059$\arcsec$ in the AL direction and 2.12$\arcsec$ in the AC direction.

In principle, each component of a faint binary with separation greater than 2.12$\arcsec$ will always be isolated in their window. Since each source will be observed multiple times with different scan directions during the 5 year \textit{Gaia} mission, the 1D data can be used to create a 2D reconstruction \citep{harrison2011}. At the current time the individual 1D scans have not been released by the \textit{Gaia} project.

\subsection{Catalogue creation}
Source measurements are based on modelling single point sources in these 2D windowed regions of pixels or 1D samples. When close pairs are encountered in the same windowed region, depending on the scanning direction and relative orientation of the pair, the fainter object is given a truncated window of pixels, which has not been processed for GDR1 \citep{fabricius2016}. Therefore there is a significant lack of detections of the fainter companion in close binaries in GDR1.

\begin{figure}
	\includegraphics[width=\columnwidth]{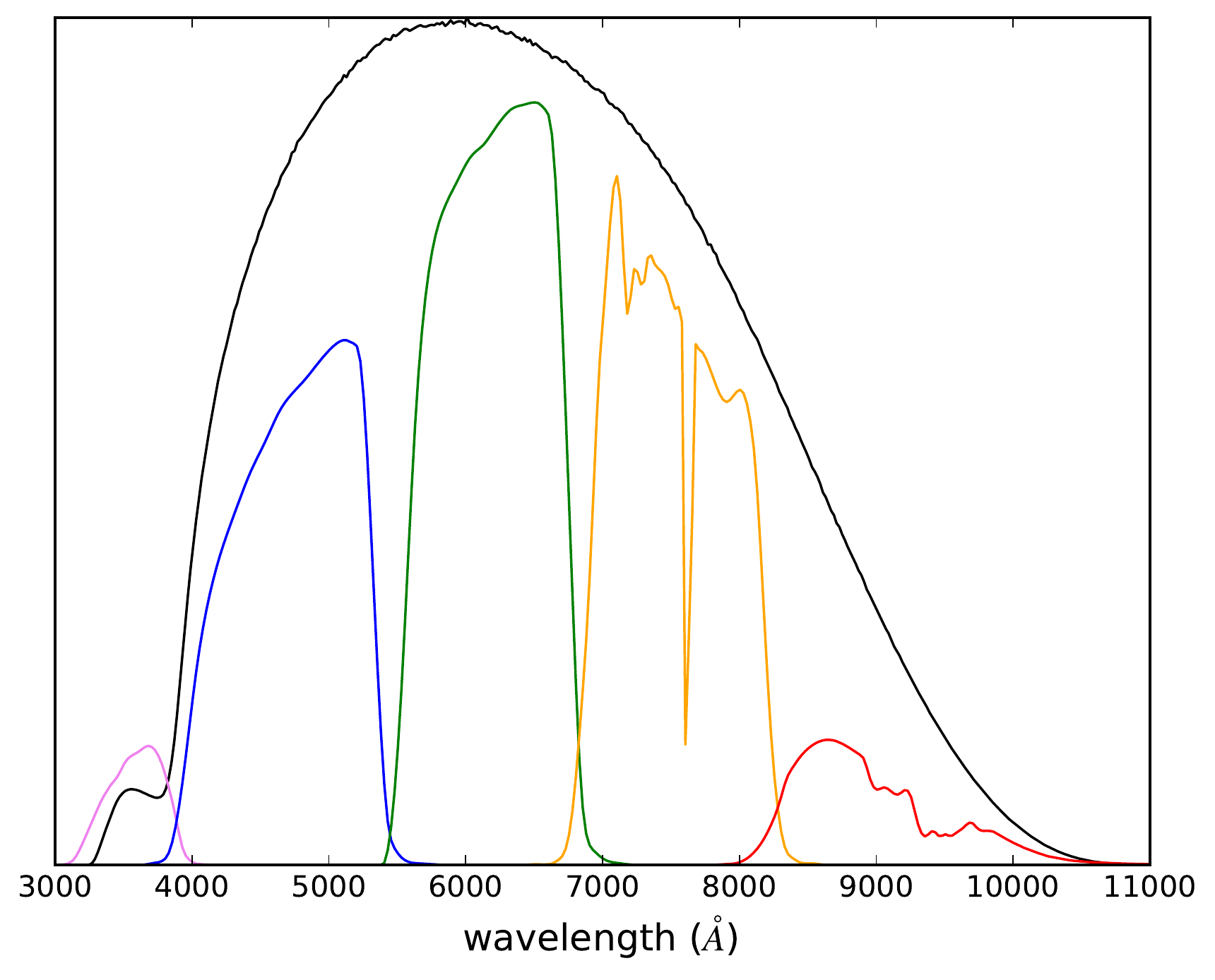}
    \caption{The nominal \textit{Gaia} G-band in black, with SDSS \textit{ugriz} filters overlaid.}
    \label{fig:Gaiagri}
\end{figure}

The main astrometric source catalogue for GDR1 contains over one billion sources (see \citet{lindgren2016} for details). The fluxes are measured in a wide optical band (hereafter G-band) as shown in Figure \ref{fig:Gaiagri}. 37 million sources were removed before GDR1 was published, when they were separated from another source by less than 5 times the combined astrometric positional uncertainty. This was due to the cataloguing of known objects against an initial \textit{Gaia} source list, which catalogued many objects twice. Further objects were filtered from GDR1 if sources were observed fewer than 5 times (5 focal plane transits), or if their astrometric excess noise and positional standard uncertainty were greater than 20mas and 100mas respectively. Finally sources were removed if they had fewer than 11 G-band measurements (CCD transits in the astrometric part of the focal plane). See \citet{fabricius2016} for a full explanation of the data-processing and catalogue creation. The limiting magnitude of \textit{Gaia} is $\sim$ 20.7 in the Vega magnitude system.

\subsection{Close pairs in \textit{Gaia}} \label{om10section}
\begin{figure}
	\includegraphics[width=\columnwidth]{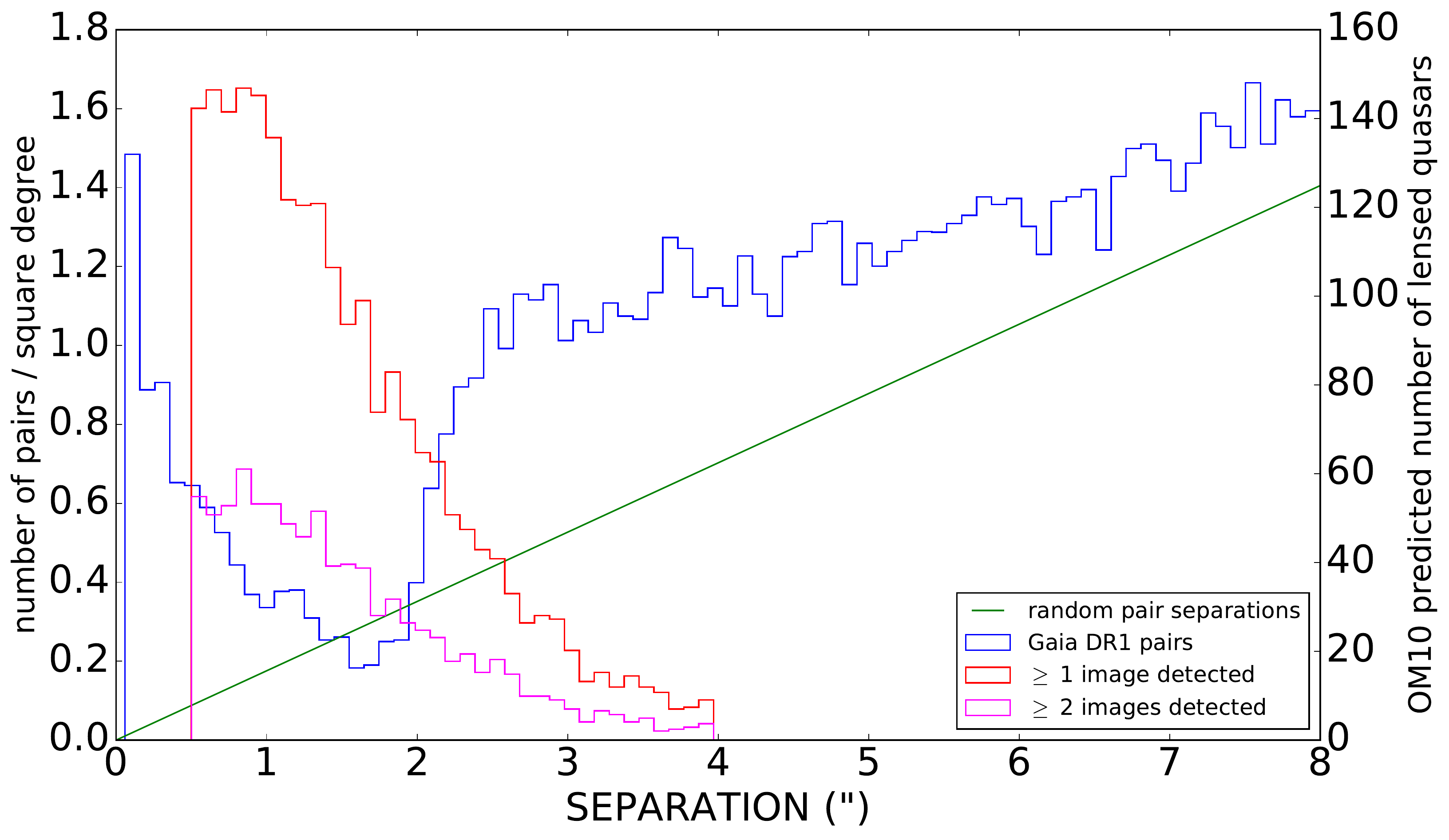}
    \caption{Expected numbers of lensed quasars as a function of image separation in 0.1$\arcsec$ bins. The plot shows expected numbers for at least two images to be detected by \textit{Gaia} (magenta) and just one image to be detected by \textit{Gaia} (red). The blue histogram is the \textit{Gaia} distribution of source pairs against separation, which shows a distinct dropout around the typical separations of lensed quasars. The green line is the distribution expected from a field of randomly-positioned sources.}
    \label{fig:om10pred}
\end{figure}

Where \textit{Gaia} can provide the most use for lensed quasar searches is in determining whether a source is composed of multiple stellar objects. In ground-based optical imaging surveys, the typically much larger FWHM causes many contaminant systems to resemble lensed quasars. These contaminants include single quasars with bright host galaxies, quasars or stars blended with galaxies, projected systems, and starburst galaxies with quasar-like colours. Because of the truncated windows given to fainter companions around brighter neighbours by \textit{Gaia}, there is a limitation in \textit{Gaia}'s handling of close pairs \citep{arenou2017}. We demonstrate how this issue affects lensed quasars in Figure \ref{fig:om10pred}. The blue histogram shows the distribution of separations of detected \textit{Gaia} pairs in a sparse field out of the galactic plane (l $\sim$ 173$^{\circ}$, b $\sim$ 67$^{\circ}$), i.e. a field in which \textit{Gaia} is able to read out all objects within the magnitude threshold. The green line is the expected distribution from randomly positioned sources on the sky matched to a large-separation asymptote. We expect an overdensity of close-separation objects relative to this line associated with stellar binaries, bright star-forming mergers and even lensed quasars, however the \textit{Gaia} catalogue significantly lacks the detection of the second object below 2.5 arcseconds. The pairs with very small \textit{Gaia} separations are possibly duplicate sources that could not be filtered from GDR1 \citep{arenou2017}. We also show the numbers of lensed quasars OM10 predicted across the whole sky with either the brightest object detectable by \textit{Gaia} (approximately equal to an i-band magnitude < 20.7) or with at least two images detectable by \textit{Gaia}, in bins of 0.1$\arcsec$ separation. The OM10 catalogue is truncated at an image separation of 0.5$\arcsec$. This cut was made due to difficulty in detection and characterisation of lenses at lower image separations, however \textit{Gaia} will likely be able to push past this limit in detection. \citet{finet2016} discuss the detection of smaller image separation lensed quasars in \textit{Gaia}, however in this paper we use the full OM10 catalogue for predicted numbers of lenses since we are interested in those that can be characterised in ground-based imaging data. The area under the magenta curve of Figure \ref{fig:om10pred} shows that later \textit{Gaia} catalogues, which will include secondary source detections in close binaries, should detect the multiple images of $\sim$ 900 lensed quasars, including $\sim$ 240 quadruply-imaged systems. This value is in agreement with other estimates based on \textit{Gaia}'s expected pre-launch capabilities \citep{finet2016, surdej2002}. The number of lensed quasars expected with only one image detected by \textit{Gaia} is $\sim$ 1400, only $\sim$ 50 of which are quads. This very low quad fraction is due to the brightest two images of quads often having similarly magnified fluxes.

\section{A sample of small-separation lensed quasars} \label{sample}

At separations larger than 2$\arcsec$, the images of lensed quasars are deblended in ground-based surveys and \textit{Gaia} cannot provide much further information, though it is still useful in some cases, for example in detecting quasar images around bright lensing galaxies, where the system might remain blended.

\begin{figure}
	\includegraphics[width=\columnwidth]{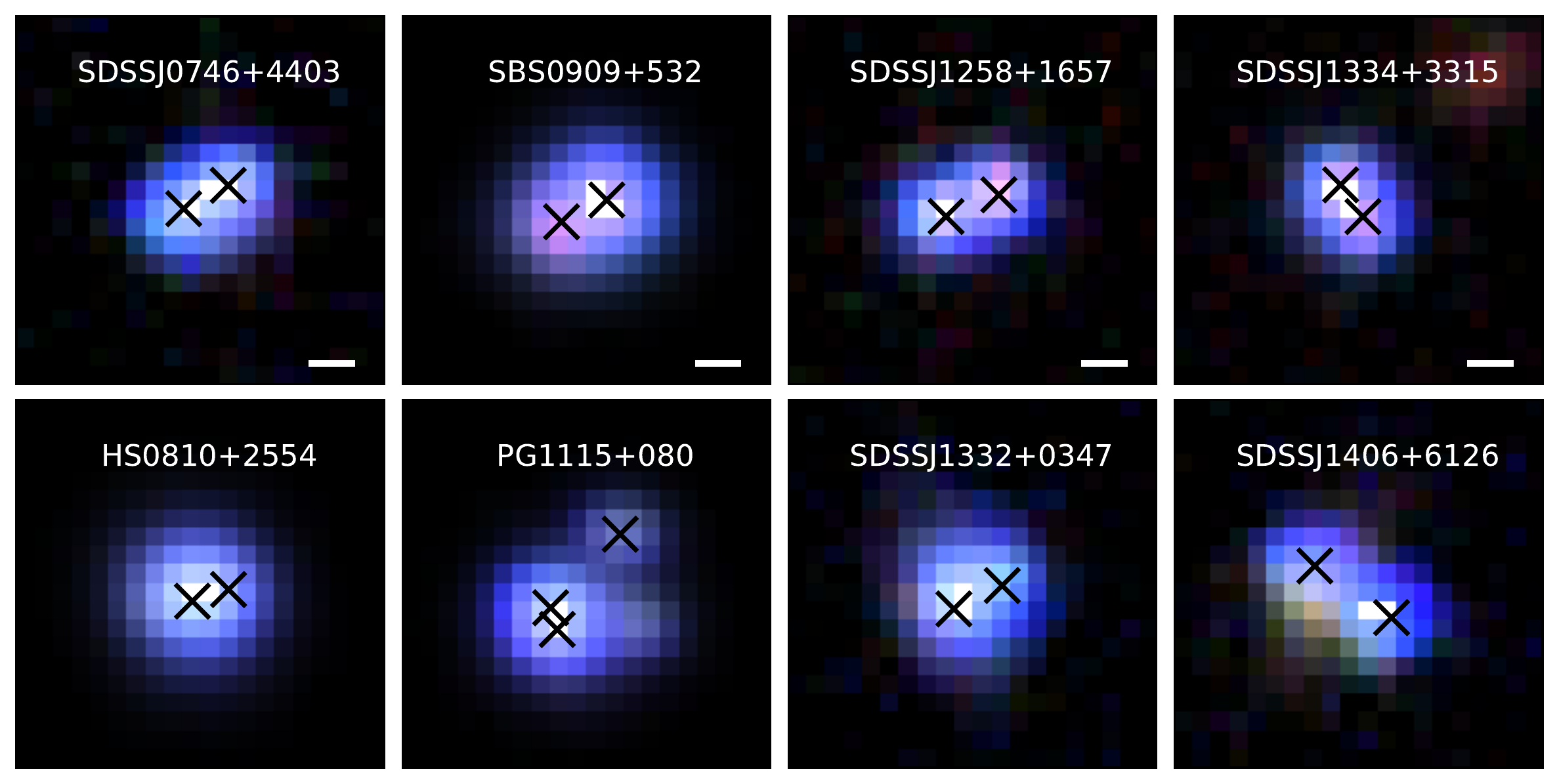}
    \caption{The 8 small-separation lensed quasars that have multiple detections in the \textit{Gaia} catalogue with positions overlaid on an SDSS \textit{gri} colour image. These lenses are: SDSSJ0746+4403 \citep{inada2007}, HS0810+2554 \citep{reimers2002}, SBS0909+532 \citep{kochanek1997}, PG1115+080 \citep{weymann1980}, SDSSJ1258+1657 \citep{inada2009}, SDSSJ1332+0347 \citep{morokuma2007}, SDSSJ1334+3315 \citep{rusu2011} and SDSSJ1406+6126 \citep{inada2007}.}
    \label{fig:Gaia_multiples}
\end{figure}

To investigate how \textit{Gaia} catalogues known close-separation lensed quasars, we compile a set of 49 lensed quasars in the SDSS footprint with image separations less than 2$\arcsec$ and which are typically blended in SDSS. All lenses in the sample have at least one \textit{Gaia} detection, but only 8 have multiple detections in GDR1, whereas 43 are expected to have further images brighter than the detection limit. Only one lens, PG1115+080, is detected as three separate images (though we note this is one of the larger-separation systems) and is deblended into 2 components in SDSS. These 8 systems are shown in Figure \ref{fig:Gaia_multiples} with \textit{Gaia} detections overlaid on SDSS \textit{gri} colour images.

Since many objects were removed as possible duplicates in GDR1, we check for duplicate removal of the non-catalogued images through the \textit{duplicated\_source} flag. Approximately 5 percent of objects in GDR1 have this flag. For the lensed quasars with single \textit{Gaia} detections where multiple detections are expected, only 4 of 35 are flagged as duplicated sources, indicating the further missing detection(s) in \textit{Gaia} may be due to truncated windows being given to the missing images in certain scan directions. Since these windows have not yet been processed, not all quasar images have enough G-band measurements to be included in GDR1 \citep{fabricius2016}.

\begin{table}
 \begin{tabular}{l|c|c|c|c|c}
  \hline
  Name & z & sep ($\arcsec$) & G & G1-G2 \\
  \hline
  SDSSJ0746+4403 & 2.00 & 1.08 & 19.39, 19.47 & 0.07 \\
  HS0810+2554 & 1.50 & 0.81 & 15.94, 16.99 & 1.05 \\
  SBS0909+532 & 1.38 & 1.09 & 16.25, 16.73 & 0.48 \\
  PG1115+080$^{a}$ & 1.72 & 0.48 & 17.08, 17.20 & 0.12 \\
  SDSSJ1258+1657 & 2.70 & 1.25 & 19.19, 19.50 & 0.31 \\
  SDSSJ1332+0347 & 1.45 & 1.15 & 19.61, 19.64 & 0.02 \\
  SDSSJ1334+3315 & 2.43 & 0.84 & 19.58, 19.80 & 0.22 \\
  SDSSJ1406+6126 & 2.13 & 1.99 & 19.77, 19.97 & 0.20 \\
  \hline
 \end{tabular}
 \caption{Details of small-separation lensed quasars with multiple \textit{Gaia} detections (source redshift, image separation and \textit{Gaia} magnitudes).  $^{a}$Values for PG1115+080 are for the close-separation merging pair.}
  \label{tab:lenspairs}
\end{table}

\section{Finding quasar lenses in \textit{Gaia}} \label{howfind}
Since small-separation lenses are much more common than the easily deblended larger-separation lenses (Figure \ref{fig:om10pred}), we now explore several ways that GDR1 can help identify these small-separation lensed quasars by considering our sample of known lenses. Firstly we consider multiple \textit{Gaia} detections corresponding to the multiple images of a lensed quasar. In the case of single \textit{Gaia} detections, we investigate the \textit{Gaia} catalogue parameter of astrometric excess noise which might hint at the presence of a lensing galaxy or further quasar images. Finally we compare \textit{Gaia} fluxes and positions to other datasets that blend lensed quasars into single objects, which will naturally measure larger fluxes and different centroids from those given in the \textit{Gaia} catalogue.

Throughout the rest of this paper we use spectroscopically confirmed objects from the \textit{SpecObjAll} table from the twelfth data release of SDSS \citep{alam2015}, in which most spectroscopically confirmed quasars are from SDSS-III BOSS, but also includes spectra from all previous SDSS data releases.

\subsection{Multiple \textit{Gaia} detections}
Our sample of small-separation lensed quasars shows that several objects are still separated into multiple components by \textit{Gaia}. \citet{arenou2017} have shown that the detection of both components of close pairs is more likely when they are of similar magnitude since the primary object can be either of the pair in different focal plane transits, leading to both objects being catalogued in GDR1. We see this in the 8 lensed quasars with multiple detections in our sample (see Table \ref{tab:lenspairs}). Therefore even in GDR1 it is possible to use \textit{Gaia}'s resolution to deblend potential quasar lens candidates into multiple objects immediately. A common limitation to finding lensed quasars in ground-based imaging surveys is the lensing galaxy causing a possible contamination to the colour of the object, meaning conventional quasar colour classification will miss these objects (see \citealp{ostrovski2016} for a full discussion). However given the prior that two point sources must be present means \textit{Gaia} can help detect these bright lens galaxy systems, relying less on constraints from quasar colour-selection techniques.

\subsection{Single \textit{Gaia} detections}
As discussed in Section \ref{om10section}, most bright lensed quasars will have one detection in GDR1. However, even when \textit{Gaia} detects a single image of a lensed quasar, it provides useful information in its catalogue values of astrometric excess noise, position and flux for each detection. The latter two measurements are useful because we are able to search for systems with missing flux and/or a different centroid relative to that of a dataset capturing all the flux, e.g. from proper imaging, where the difference would be caused by \textit{Gaia} not cataloguing all sources. That is, GDR1 effectively resolves out the flux from single images of lensed quasars. For a given object, a potential close-separation quasar lens or perhaps just a single quasar, we outline how to derive a synthesised \textit{Gaia} magnitude from ground-based broad-band survey photometry, with which one can compare the \textit{Gaia} magnitude. In the case of a large discrepancy, this could be accounted for by the presence of extra quasar images and a lensing galaxy, which the ground-based imaging has blended with the original \textit{Gaia} detection. To determine the centroid difference we use SDSS positions, which have been recalibrated as in \citet{deason2017}. The detection of only one image of a lensed quasar in \textit{Gaia} will not be limited to the first \textit{Gaia} data release. Indeed $\sim$ 1400 lenses (Section \ref{om10section}) will have just one component bright enough to be detected by \textit{Gaia} but have other images fainter than the magnitude threshold, as can be seen by the difference between the two lensed quasar population histograms in Figure \ref{fig:om10pred}.

\subsubsection{Astrometric Excess Noise}
\begin{figure}
	\includegraphics[width=\columnwidth]{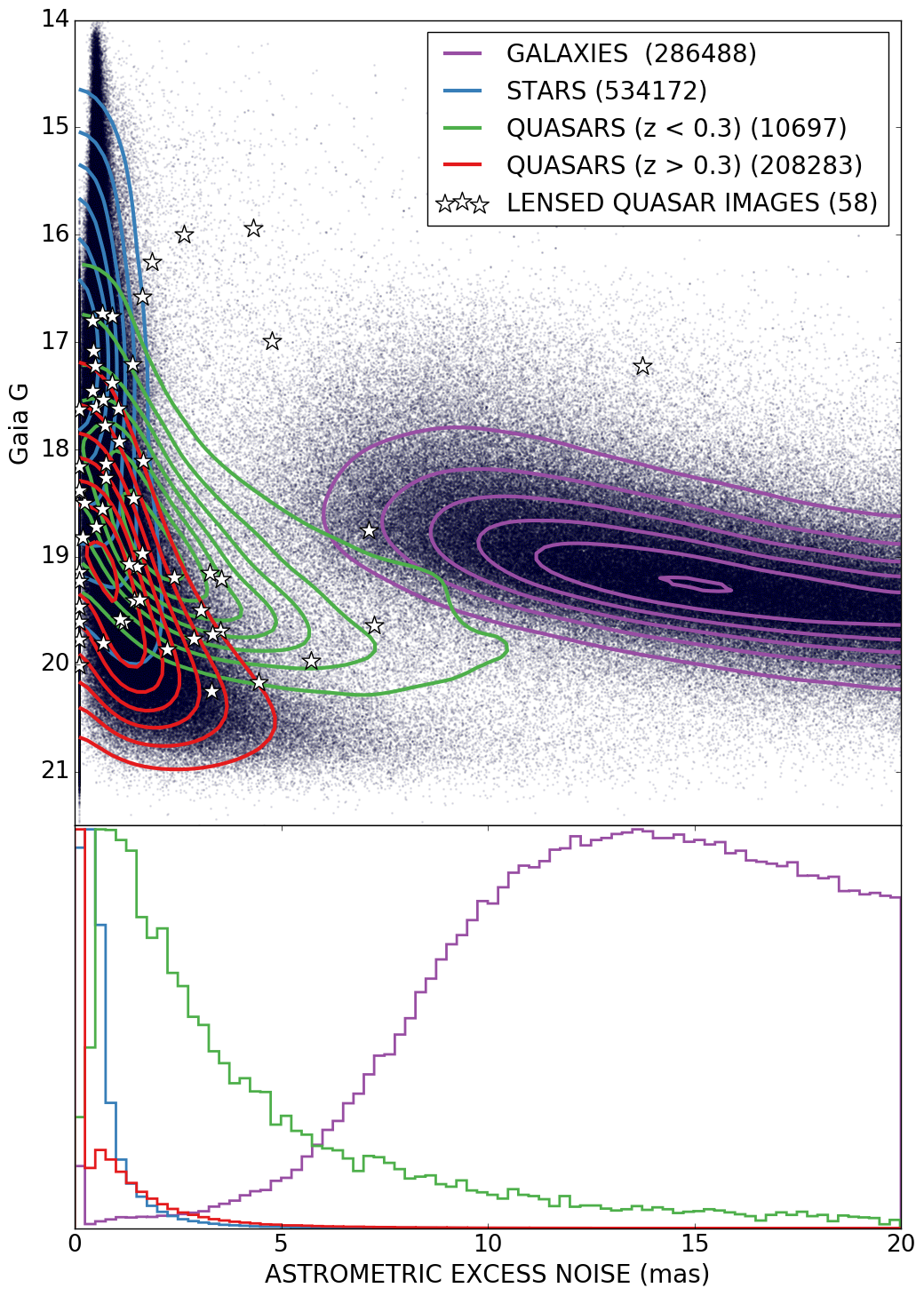}
    \caption{Distributions of astrometric excess noise and \textit{Gaia} magnitudes for spectroscopically confirmed galaxies, stars and quasars matched to \textit{Gaia} with all lensed quasar images from our small separation sample overlaid}
    \label{fig:Gaia_aen}
\end{figure}

We initially consider the catalogue value of \textit{astrometric excess noise} (hereafter AEN). This parameter represents the scatter in the astrometric model for a single object \citep{lindegren2012, lindgren2016}. A large AEN might indicate the presence of uncatalogued, nearby images in lenses or perhaps a bright lensing galaxy. We match \textit{Gaia} observations to the SDSS spectroscopic catalogues for stars, galaxies and quasars within 0.5$\arcsec$. We further require no flags from the spectral classification and recover 534172 stars, 286488 galaxies and 218980 quasars. We plot their magnitude and AEN distributions in Figure \ref{fig:Gaia_aen}, splitting the quasar sample into those above and below redshift 0.3. This plot reflects the method to separate galaxies from stars using AEN, as used in \citet{belokurov2016, koposov2017}. From the overlaid positions corresponding to lensed quasars, AEN alone cannot sufficiently identify most lensed quasars, however it is a useful parameter to consider for the brightest candidates. Therefore we turn our attention to using other datasets and \textit{Gaia} catalogue parameters to indicate missing components from the catalogue.

\subsubsection{Flux deficit}\label{fluxdefssection}
\textit{Gaia}'s 0.1$\arcsec$ PSF FWHM means that it can completely resolve the images of any useful lensed quasar (those that have large enough image separation to be spectroscopically confirmed). Therefore the \textit{Gaia} flux measurement is truly an indication of the magnitude of a single image of the system. Comparing this value to a ground-based flux measurement of the entire blended system may allow the detection of the presence of further components in the system. Firstly we must determine a relationship between the \textit{Gaia} G-band and ground-based photometric magnitudes.

To derive an empirical G-band fit for quasars, we need an isolated sample of quasars with \textit{Gaia} detections. We use the extensive spectroscopically confirmed quasar catalogues from SDSS, which we match to the \textit{Gaia} secondary source catalogue. On attempting to fit an empirical relation between the SDSS \textit{ugriz} magnitudes and \textit{Gaia}, the variability over the  $\sim$ 10 year mean epoch difference is apparent (2015 for \textit{Gaia} and 2003 for SDSS). Therefore we use the much better matched mean epoch of Pan-STARRS \citep{chambers2016}, 2013.

For the photometric fit we use the Pan-STARRS PSF magnitudes to ensure that we are comparing to the flux from just the quasar, since this is what \textit{Gaia} is measuring. We apply the following cuts to the combined spectroscopic and photometric catalogues:

\begin{enumerate}
\item spectroscopic class = QSO
\item z < 5
\item zerr < 0.05
\item zwarning = 0

\item no \textit{Gaia}, SDSS or Pan-STARRS neighbours within 5$\arcsec$
\item type = 6 (PSF morphology in SDSS)
\item Pan-STARRS $r_{\rm PSF}$-$r_{\rm kron}$ < 0.05
\item \textit{Gaia}-SDSS centroid distance < 0.1$\arcsec$
\item \textit{Gaia}-SDSS proper motion < 5mas yr$^{-1}$
\end{enumerate}

Though the SDSS PSF morphology removes the majority of objects with bright hosts or nearby neighbours, we also apply an $r_{\rm PSF}$-$r_{\rm kron}$ magnitude cut to the deeper Pan-STARRS data to remove further extended objects. Finally once an empirical fit to the G-band photometry is made, we remove those with the most inconsistent flux when compared to \textit{Gaia} ($5 \sigma$ outliers in a given magnitude bin) and repeat the fit. We are left with a catalogue of 117599 \textit{Gaia}-matched isolated quasars. 

We are now able to fit an empirical G-band magnitude from the Pan-STARRS photometry, via a simple linear combination of g, r, i and z:

\begin{ceqn}
\begin{equation}
	\rm{G_{synth}} = \alpha + \rm{r} + \beta (\rm{g-r}) + \gamma (\rm{r-i}) + \delta (\rm{i-z})
\label{eq:gri1}
\end{equation}
\end{ceqn}

\begin{ceqn}
\begin{equation}
	{\sigma}^{2} = {\sigma_{int}}^{2}  + {\sigma_{G}}^{2} + {\sigma_{Gsynth}}^{2}
\label{eq:gri2}
\end{equation}
\end{ceqn}

\begin{ceqn}
\begin{equation}
	\rm{log} P = -\frac{\left({G-G_{synth}}\right)^{2}}{2 {\sigma}^{2}}-\frac{\rm{log}(2 \pi {\sigma}^{2})}{2}
\label{eq:gri3}
\end{equation}
\end{ceqn}

 This is well-motivated since the nominal \textit{Gaia} bandpass roughly overlaps these optical filters, as shown in Figure \ref{fig:Gaiagri}. To include the high-redshift quasars in the sample we allow for their g-band dropout by fitting for two separate formulae at some g-r cut. We optimise using the log likelihood of equation \ref{eq:gri3}, determining the parameters $\alpha, \beta, \gamma, \delta, \sigma_{int}$ and the g-r cutoff.

\begin{figure}
	\includegraphics[width=\columnwidth]{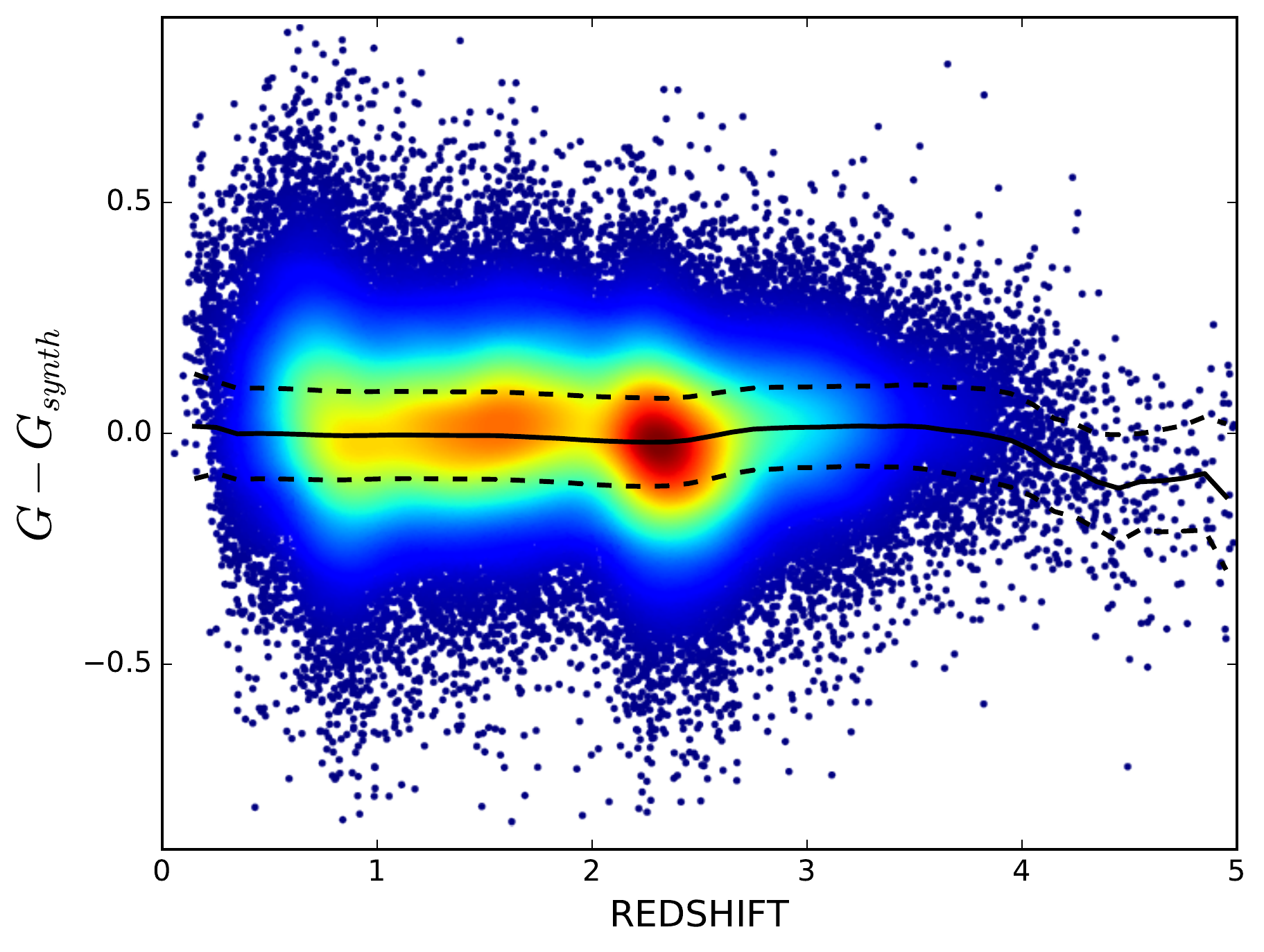}
	\hspace*{0.65cm}\includegraphics[width=0.93\columnwidth]{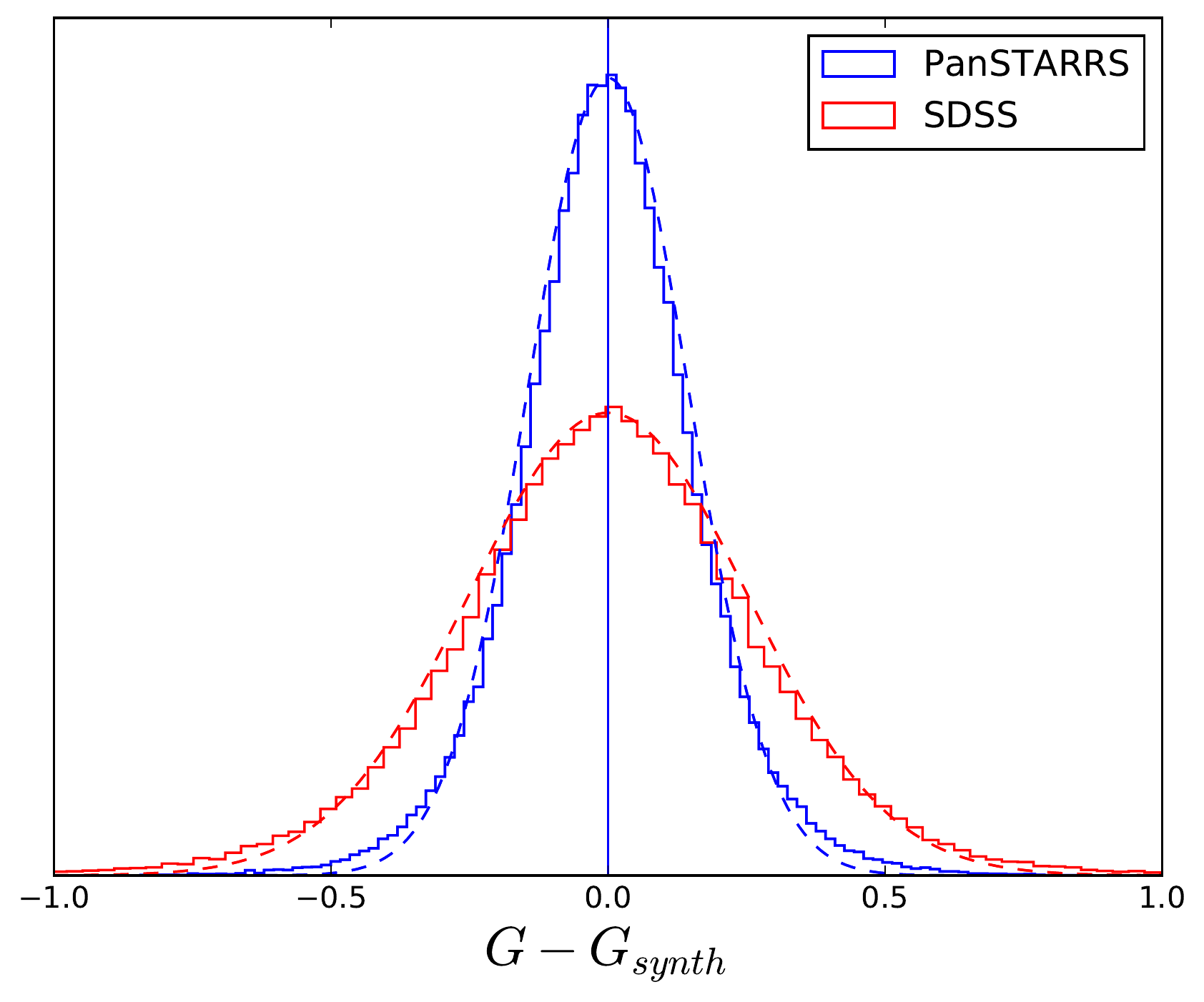}
    \caption{The top plot shows synthesised \textit{Gaia} G-band residuals against redshift for our sample of 117599 isolated quasars. The overdensities are due to the non-uniform redshift distribution. The median and median absolute deviations are plotted at each redshift (black solid and dotted lines respectively). The bottom plot compares the fit to SDSS and Pan-STARRS data and equivalent Gaussians from the median absolute deviation are overlaid. }
    \label{fig:PS_grifit}
\end{figure}

On applying this method to SDSS DR9 photometry we find an intrinsic dispersion of 0.263 mag. The fit to Pan-STARRS gives an improved intrinsic dispersion of 0.151 mag, reflecting the increased brightness variation of quasars over larger timescales. The best fit formulae using Pan-STARRS photometry\footnote{The best fit formulae using SDSS photometry are:
\begin{align*}
\rm{g-r} < 0.24: \rm{G} &=  0.029 + \rm{r} + 0.139 (\rm{g-r}) - 0.641 (\rm{r-i}) - 0.496 (\rm{i-z}) \\
\rm{g-r} > 0.24: \rm{G} &= -0.038 + \rm{r} + 0.138 (\rm{g-r}) - 0.400 (\rm{r-i}) - 0.163 (\rm{i-z}) 
\end{align*}
} are:

\begin{flalign}
& \rm{g-r} < 0.11: & \nonumber \\ 
& \rm{G} = -0.033 + \rm{r} + 0.131 (\rm{g-r}) - 0.660 (\rm{r-i}) - 0.162 (\rm{i-z}) 
\end{flalign}
\label{eq:ps_fit1}

\begin{flalign}
& \rm{g-r} > 0.11: & \nonumber \\
& \rm{G} = -0.061 + \rm{r} + 0.217 (\rm{g-r}) - 0.548 (\rm{r-i}) - 0.013 (\rm{i-z})
\end{flalign}
\label{eq:ps_fit2}

Figure \ref{fig:PS_grifit} shows the residual magnitudes against redshift. The non-uniform redshift distribution of the quasar sample does not significantly bias the fit given the very similar dispersion at all redshifts. Furthermore the residuals are essentially Gaussian-distributed about zero (Figure \ref{fig:PS_grifit}). Estimates of median variability for this sample between \textit{Gaia} and Pan-STARRS or SDSS are $\sim$0.1 or $\sim$0.15 respectively \citep{hook1994, macleod2012}. Our dispersions are larger than these estimates because of the error from the simple model fit to all quasar spectral shapes and redshifts.

We apply these empirical G-band fits to the sample of lensed quasars and compare them to \textit{Gaia} measurements in Section \ref{lenscandselect}. 

\subsubsection{Positional Offsets}\label{posoffsetssection}
Quasar lenses are often discovered by looking for extended quasar candidates. This is done by performing a cut in a PSF magnitude minus a model magnitude (fit to a galaxy profile), where it is expected that a PSF magnitude will not capture all the flux of an extended object. However this technique may fail to find all lensed quasars, especially those with small separations. To determine the limitations of this method, we match all \textit{Gaia} pairs (two catalogue detections) between 0.5$\arcsec$ and 1.0$\arcsec$ separation to SDSS and compute $r_{\rm PSF}$-$r_{\rm model}$ \citep[in this case we use the SDSS PSF and CMODEL magnitudes in the r-band][]{stoughton2002}. We keep all pairs with brightness differences within 1 magnitude (based on their \textit{Gaia} G-band magnitudes) as is typical for lensed quasar image fluxes, and remove objects obviously made up from non-point sources. We find a spread in $r_{\rm PSF}$-$r_{\rm model}$ of $0.076 \pm 0.049$ for objects with image separation 0.5-0.6$\arcsec$ and $0.215 \pm 0.117$ for those with image separation 0.9-1.0$\arcsec$. To understand these values we compare to single PSF objects (isolated \textit{Gaia} detections) and find a value of $0.003 \pm 0.042$. Therefore, given the overlap in the $r_{\rm PSF}$-$r_{\rm model}$ values for single stars and pairs at separations of 0.5$\arcsec$, the $r_{\rm PSF}$-$r_{\rm model}$ magnitude comparison for classification of extended/point source objects will often not be able to distinguish between relatively close binaries and single stellar objects. Furthermore, the sample of pairs we have used here is highly biased to objects of similar magnitude (as explained in \citealp{arenou2017} and due to our magnitude difference cut) whereas objects with large flux ratios will tend to appear even more PSF-like, implying a conservative upper bound on our $r_{\rm PSF}$-$r_{\rm model}$ values for pairs.

Given this discussion, it is clear that the PSF classification of SDSS ($r_{\rm PSF}$-$r_{\rm model}$ < 0.145) will class many binaries and small separation lensed quasars as PSFs. This is perhaps less likely for quad lensed quasars in which there are four images; however, the image separations of the brightest components can often be much smaller than the Einstein radius, leading to a more PSF-like appearance. 

Therefore we again turn to \textit{Gaia}'s excellent resolution to define a parameter indicative of multiple system components. When a system is composed of two close stellar components, \textit{Gaia} will catalogue the centroid of one of the two components with high precision and accuracy. However the same system in ground-based imaging will be blended and the catalogued centroid will lie between the two objects, offset from the \textit{Gaia} centroid. This offset can easily be calculated and should be large for crowded systems like lensed quasars. \citet{jackson2007} suggest searching for an offset between optical and radio positions for identifying lensed quasars in radio surveys. Our method is similar however we rely on both the galaxy and uncatalogued quasar images to cause the offset and only require optical data. While this offset might not be significant for doubles with large flux ratios, we always expect it to be apparent for quadruply-lensed quasars, since flux ratios between the two brightest images are approximately unity.

To robustly compare the difference in \textit{Gaia} and SDSS positions, we must understand the minimum uncertainty expected from fitting a single PSF. Using a Gaussian PSF defined only by the FWHM and assuming at least critical sampling, any unbiased estimator is limited to a centroid positional error of $\sim$ (FWHM)/(signal to noise) \citep{mendez2013}. The FWHM, signal and noise are inferred from the SDSS catalogues in the r-band since the standard SDSS positions are derived from the r-band. In order to look for sub-pixel offsets we must have very accurate SDSS astrometry. We therefore use the \textit{Gaia}-based calibration of SDSS positions explained in \citet{deason2017}. Using this catalogue, the median \textit{Gaia}-SDSS offset for quasars brighter than 19th magnitude is 0.02$\arcsec$. We use these recalibrated SDSS positions to define an offset parameter (OP):

\begin{ceqn}
\begin{equation}
\text{OP} = \frac{\text{distance between \textit{Gaia} and SDSS centroids}}{\text{SDSS PSF centroid uncertainty}}
\end{equation}
\end{ceqn}

\begin{figure}
	\includegraphics[width=\columnwidth]{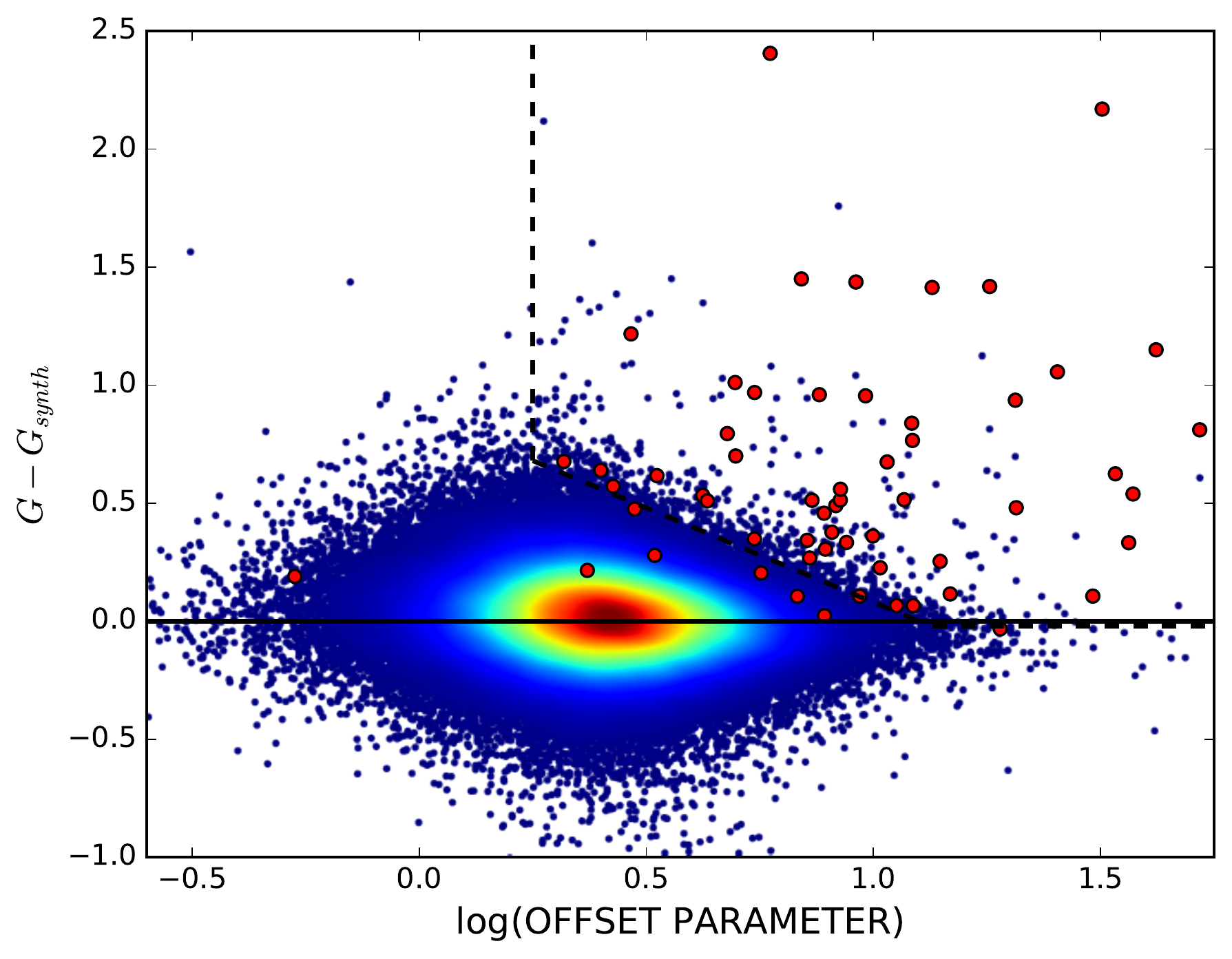}
    \caption{Spectroscopically confirmed quasars (blue) and known lensed quasars (red) plotted in the parameter space described in the text. The upper-right region indicates an area of centroid and flux disagreement between \textit{Gaia} and ground-based optical data that contains a large number of lensed quasars and few single quasars.}
    \label{fig:offsets_fluxdef}
\end{figure}

The spread of this offset parameter in combination with the flux deficit from Section \ref{fluxdefssection} can be seen for lenses and SDSS spectroscopically confirmed quasars in Figure \ref{fig:offsets_fluxdef} (see Section \ref{singledetqsos} for details).

This plot confirms the intuition that lensed quasars should have centroid and flux differences between \textit{Gaia} and ground-based survey measurements. We use this plot in the next section to search for new lens candidates.

\section{Lens candidate selection} \label{lenscandselect}
GDR1 gives several clues to a source's nature, but cannot be used to distinguish between quasars and stars alone. Therefore we must start from a quasar candidate catalogue to find lensed quasars. The results of a search for lensed quasars using  \textit{Gaia} detections in a variety of quasar catalogues are presented in a companion paper (Lemon et al. in prep.). However to better understand the limitations of our methods, we will only consider the spectroscopically confirmed quasars of SDSS as a starting catalogue in this paper. The extra information from the spectra allows us to identify contaminant systems.

Many of the SDSS spectroscopically confirmed quasars have already been targeted by extensive lens follow-up programmes, namely SQLS and the BOSS quasar lens search, both of which have confirmed the identity of many lenses and other contaminant close-separation systems \citep{inada2008, inada2010, inada2012, more2016}. However these campaigns focussed on lenses with larger image separations than 1$\arcsec$, so we expect lenses still exist in this catalogue at smaller separations. We restrict the redshift range of the SDSS quasars to 0.6 < z < 5 to avoid low-redshift host galaxies creating outliers in positional and flux offsets and incorrectly identified objects as high-redshift quasars. Upon matching this set of quasars to \textit{Gaia} detections within 3$\arcsec$, we find 199376 objects which we take as our catalogue for the following searches.

\subsection{Multiple detections}
We search the quasar catalogue for multiple \textit{Gaia} detections within 1.5$\arcsec$ to target the lenses that might have been missed by previous searches.  We find that only 74 systems from our quasar catalogue have multiple detections in \textit{Gaia} up to 1.5$\arcsec$ and visually inspect their Pan-STARRS images. 7 of the 74 systems are the lensed quasars in our original sample (Section \ref{sample}). Another 7 objects were falsely classified as high-redshift quasars. Approximately half of these 74 systems have very small separations (< 0.25$\arcsec$) and appear as single PSFs in Pan-STARRS. Of the remaining higher-separation candidates, they are either obvious quasar and star pairs (indicated by large colour differences or stellar spectral features in the quasar spectra) or have already been followed up by SQLS and identified to be binary quasars or potential small-separation lenses.

\subsection{Single detections} \label{singledetqsos}
We calculate the flux deficit and offset parameter for our quasar catalogue and lens sample. The Pan-STARRS magnitudes are used to calculate a synthetic G-band magnitude as described in \ref{fluxdefssection}. However since we want the total flux from the ground-based survey instead of the PSF magnitudes we used to fit the relation, we use the KRON magnitudes. The parameter values are plotted in Figure \ref{fig:offsets_fluxdef} for both the quasar and lensed quasar samples. As expected the lenses populate the area in which \textit{Gaia} and SDSS have disagreeing centroids, and where \textit{Gaia} is missing flux relative to Pan-STARRS. We note that the outlier with the smallest offset is SDSSJ0737+4825 \citep{more2016}. This system is a faint double with a large flux ratio (i-band magnitudes of 18.28 and 20.58, a flux ratio of $\sim$ 8.5). Therefore, as we expected, it does not have a large statistical offset to the \textit{Gaia} detection. Furthermore the other lenses that lie towards the bottom left of the plot are doubles. All quadruply-imaged lenses are well-separated from the single quasars since the extra images cause larger flux deficits and more robust offsets, as already predicted in Section \ref{posoffsetssection}.

Based on the offset parameters (OP) and flux deficits (FD), $G-G_{synth}$, of our lens sample (Figure \ref{fig:offsets_fluxdef}) we search for possible new lenses in our quasar catalogue by inspecting quasars with similar parameter values to the known lenses. We define a region in Figure \ref{fig:offsets_fluxdef} which clearly retains the majority of lenses while including very few of the main spectroscopic sample. The region is defined as log(OP) > 0.25, FD > $-0.05$ and FD+0.86log(OP) > 0.914, and contains 362 objects in the quasar catalogue (from the original sample of 199376). We find about 10 percent of these objects are associated with blue stars with featureless spectra that the SDSS pipeline has classified as quasars without any spectrum-based warning. These objects have large OP and FD values perhaps due to proper motions and the empirical G-band values being based on a fit to quasars. After removing these from our sample by requiring a WISE detection \citep{wright2010}, we find 319 possibly-extended candidates, of which 63 have Canada France Hawaii Telescope (CFHT) archival data or Hyper Suprime Cam \citep[HSC,][]{aihara2017}) Survey data, which we inspect. Many do not show distinct components either due to being outliers from the single quasar population (e.g. highly-variable quasars between Pan-STARRS and \textit{Gaia} measurements) or having component separations small enough to become blended in the ground-based imaging.

\begin{figure}
	\includegraphics[width=\columnwidth]{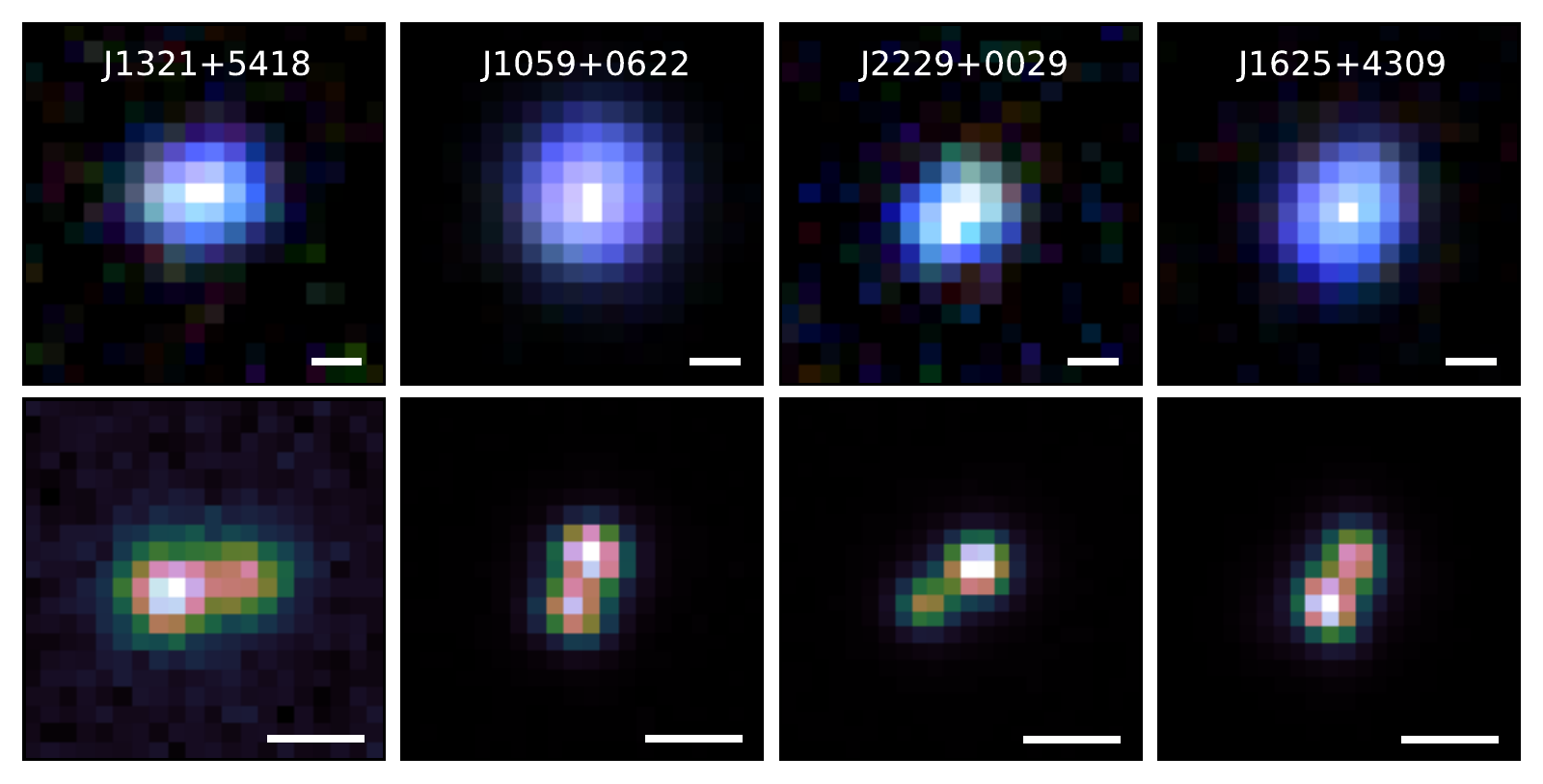}
    \caption{New SDSS lensed quasar candidates. The top images show SDSS \textit{gri} cutouts and bottom are either CFHT r-band cutouts (J1321+5418, J1059+6200) or HSC i-band cutouts (J2229+0029, J1625+4309) in a cubehelix colour scheme \citep{green2011}. The white bar is scaled to 1$\arcsec$. Details for these systems can be found in Table \ref{tab:candidatepairs}.}
    \label{fig:cfht_singleGaia}
\end{figure}

\begin{figure}
	\includegraphics[width=\columnwidth]{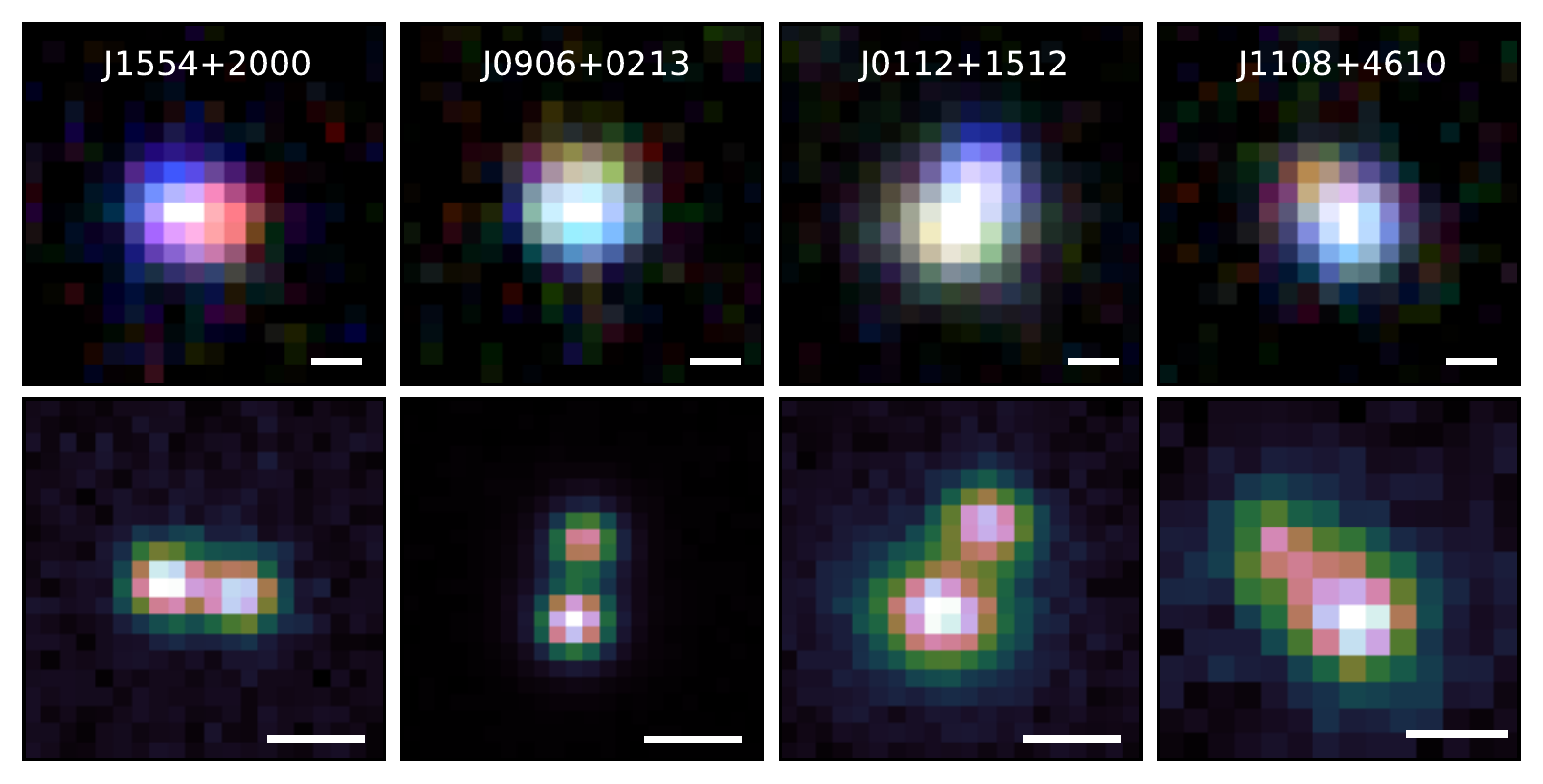}
    \caption{SDSS quasar contaminant systems. The top images show SDSS \textit{gri} cutouts and bottom are either CFHT r-band (J1554+2000, J0112+1512), HSC i-band (J0906+0213) or Pan-STARRS z band (J1108+4610). The white bar is scaled to 1$\arcsec$. Details for these systems can be found in Table \ref{tab:candidatepairs}.}
    \label{fig:cfht_singleGaia_contams}
\end{figure}

We find 4 objects with clearly resolved components that have similar colours and do not show obvious contaminant spectral features (e.g. stellar absorption lines). These are shown in Figure \ref{fig:cfht_singleGaia} and several properties for the systems are listed in Table \ref{tab:candidatepairs}. Examples of the contaminant systems that can be distinguished through their spectra or colour images are shown in Figure \ref{fig:cfht_singleGaia_contams} and their properties are also included in Table \ref{tab:candidatepairs}. Each of these systems was classed as a contaminant because of a strong colour gradient and, in the case of J0112+1512, the presence of an extended source. While we might be seeing lensing galaxies, no other quasar image is apparent and such large colour differences are unlikely between quasar images. These detections further demonstrate the effectiveness of our method in selecting close-separation pairs.

\begin{table*}
 \caption{Details of new candidate lensed quasars (LC) and contaminant systems (CS).}
 \label{tab:candidatepairs}
 \begin{tabular}{cccccccccc}
  \hline
  Name & classification & RA & DEC & redshift & separation($\arcsec$) & $G$ & $G-G_{synth}$ & centroid offset($\arcsec$) & log(OP)\\
  \hline
  J1321+5418 & LC & 200.36941 & 54.31544 & 2.26 & 0.81 & 20.24 & 0.429 & 0.226 & 0.74 \\
  J1059+0622 & LC & 164.86015 & 6.37420 & 2.19 & 0.59 & 17.48 & 0.337 & 0.253 & 1.31 \\
  J2229+0029 & LC & 337.42847 & 0.49840 & 2.14 & 0.64 & 20.48 & 0.523 & 0.244 & 0.70 \\
  J1625+4309 & LC & 246.25829 & 43.15872 & 1.65 & 0.53 & 19.27 & 0.345 & 0.231 & 0.94 \\
  J1554+2000 & CS & 238.58140 & 20.00245 & 2.25 & 0.73 & 20.38 & 0.499 & 0.426 & 0.84 \\
  J0906+0213 & CS & 136.72724 & 2.22088 & 2.01 & 0.82 & 20.35 & 0.769 & 0.060 & 0.42 \\
  J0112+1512 & CS & 18.12246 & 15.20406 & 1.96 & 0.99 & 19.80 & 0.479 & 0.712 & 1.07 \\
  J1108+4610 & CS & 167.18778 & 46.17658 & 1.84 & 0.97 & 19.99 & 0.494 & 0.117 & 0.64 \\
  \hline
 \end{tabular}
\end{table*}

\section{Conclusions}
Given the sky coverage, depth and excellent resolution of \textit{Gaia}'s first data release, it is a prime dataset to use for lensed quasar searches. Starting from a quasar candidate catalogue, the \textit{Gaia} source catalogue can be searched for multiple detections for each candidate, quickly identifying likely lensed quasars. This method will identify $\sim$900 lensed quasars ($\sim$240 quads) with image separations above 0.5$\arcsec$ across the whole sky \citep{oguri2010, finet2016}. We perform this search in a variety of photometric quasar catalogues and present the results in a companion paper (Lemon et al. in prep).

However, GDR1 has often catalogued only one component of lensed quasars, in particular for systems with small image separations \citep{arenou2017}. It is at these separations that we would benefit most from \textit{Gaia}, since ground-based imaging is unable to resolve the separate images of a lensed quasar. Therefore we have developed a method to exploit a single \textit{Gaia} detection to find the population of small-separation lensed quasars. This method relies on \textit{Gaia} effectively resolving out the flux from just one image of a lensed quasar, whereas ground-based observations (from SDSS and Pan-STARRS) blend the components together. For a sample of 49 known small-separation lensed quasars, we demonstrate that the Pan-STARRS flux is larger than the \textit{Gaia} flux, verifying the idea that \textit{Gaia} is only measuring a single image. We also show that the offset between \textit{Gaia} and SDSS positions for our sample of lensed quasars is significant, because the \textit{Gaia} centroid lies on top of the detected image, whereas in SDSS it is at the luminosity-weighted centroid of the system.

By performing the same flux and centroid difference measurements on spectroscopically confirmed SDSS quasars, we are able to search for new lensed quasars. Inspecting better-seeing data of the systems with the largest flux and centroid offsets, we find 4 new sub-arcsecond lensed quasar candidates with resolved components in HSC or CFHT data. At such small separations, projected systems (e.g. quasar+star) are less common, and so lensed quasars selected in this way are less contaminated.

As future \textit{Gaia} data releases improve the completeness of secondary source detection in close pairs, multiple \textit{Gaia} detections will become an easily-implemented method to find lensed quasar candidates. However, our method of using single \textit{Gaia} detections will still be a useful tool to discover lensed quasars that have only one image bright enough to be detected by \textit{Gaia}. This will include $\sim$1400 lensed quasars. Furthermore, as wide-field optical surveys become deeper, the centroid offset will become a more robust statistic for differentiating between single quasars and lensed quasars. The task will then be to remove contaminant systems such as quasar and star alignments. \textit{Gaia}'s long temporal baseline and repeated measurements will help select systems with similar component variability and its blue and red photometer instruments can ensure components have a similar colour.

Finally we note that these methods are not restricted to lensed quasar searches, but should be useful for searches for stellar binary companions, or to remove contaminants from samples of isolated quasars. These techniques demonstrate how \textit{Gaia}'s excellent resolution provides an important complement to future deep ground-based optical surveys.

\section*{Acknowledgements}

We thank Diana Harrison and Lindsay Oldham for useful discussions and comments about the paper. CAL would like to thank the Science and Technology Facilities Council (STFC) for their studentship. MWA also acknowledges support from the STFC in the form of an Ernest Rutherford Fellowship. This paper includes data based on observations obtained at the Canada-France-Hawaii Telescope (CFHT) which is operated by the National Research Council of Canada, the Institut National des Sciences de l'Univers of the Centre National de la Recherche Scientifique of France, and the University of Hawaii.




\bibliographystyle{mnras}
\bibliography{papers}

%



\appendix

\bsp	
\label{lastpage}
\end{document}